\documentclass{article}

\usepackage{float}
\usepackage{PRIMEarxiv}
\usepackage{amsmath}
\usepackage[utf8]{inputenc} 
\usepackage[T1]{fontenc}    
\usepackage{hyperref}       
\usepackage{url}            
\usepackage{booktabs}       
\usepackage{amsfonts}       
\usepackage{nicefrac}       
\usepackage{microtype}      
\usepackage{lipsum}
\usepackage{fancyhdr}       
\usepackage{graphicx}       
\graphicspath{{media/}}     

\title{RAG-Fusion: a New Take on Retrieval-Augmented Generation
}

\author{
  Zackary Rackauckas \\
  Infineon Technologies \\
  San Jose, CA\\
  \texttt{zackary.rackauckas@infineon.com} \\
}

\begin{document}
\maketitle

\begin{abstract}
Infineon has identified a need for engineers, account managers, and customers to rapidly obtain product information. This problem is traditionally addressed with retrieval-augmented generation (RAG) chatbots, but in this study, I evaluated the use of the newly popularized RAG-Fusion method. RAG-Fusion combines RAG and reciprocal rank fusion (RRF) by generating multiple queries, reranking them with reciprocal scores and fusing the documents and scores. Through manually evaluating answers on accuracy, relevance, and comprehensiveness, I found that RAG-Fusion was able to provide accurate and comprehensive answers due to the generated queries contextualizing the original query from various perspectives. However, some answers strayed off topic when the generated queries' relevance to the original query is insufficient. This research marks significant progress in artificial intelligence (AI) and natural language processing (NLP) applications and demonstrates transformations in a global and multi-industry context.
\end{abstract}


\section*{Introduction}
Infineon account managers and field application engineers have expressed the need to obtain sales-oriented product information rapidly, but Infineon's product selection guides and datasheets are often hundreds of pages long. Born from this need was the rapid development of chatbots to provide account managers and engineers with technical information in seconds. These chatbots are built off the state-of-the-art retrieval-augmented generation framework.

Recently, retrieval-augmented generation (RAG) has been at the core of all of Infineon's virtual assistants. Retrieval-augmented generation answers a user's questions related to the purpose of the virtual assistant. The method combines text generation from large language models (LLMs) and document retrieval from databases of related documents to generate accurate, relevant, and comprehensive responses. (Yu 2022) Large language models are advanced natural language processing systems trained on large sets of data that process and generate text. While they are designed to handle tasks like machine translation, summarization, and conversational interactions, RAG models can also leverage them for information retrieval. \cite{LLMs} RAG has shown remarkable success in several knowledge-intensive natural language processing tasks including accurate trivia question answering, highly accurate fact verification, and accurately answering middle school math questions. \cite{MATH} \cite{RAGK} In addition, retrieval-augmented generation reduces misinformation typically produced by large language models and non-RAG chatbots. \cite{RAGHALL}

Document retrieval is a fundamental component of the algorithm. Traditional RAG virtual assistants rank documents in the order of relevance to the query, usually by vector distances. This means that the more relevant a document is in a query, the higher priority it takes being in the answer. Recently, however, developers and researchers have explored implementing different reranking methods for documents. It has been found that reranking in retrieval-augmented generation plays a significant role in improving retrieval results and in increasing the accuracy, relevance, and comprehensiveness of answers. \cite{RAGR}

Reciprocal rank fusion (RRF) is a reranking algorithm that gives a reciprocal rank to documents in multiple sources, then combines those ranks and documents into one final reranked list. It has been found that RRF outperforms many other document reranking methods while being sensitive to its parameters. \cite{RRO} \cite{RRA} Utilizing RRF as a reranker in a RAG algorithm yields RAG-Fusion, a novel RAG-based chatbot model popularized by Adrian H. Raudaschl. \cite{RAGF}

The chatbot I developed in this paper was initially a traditional RAG bot developed to be used by automotive field engineers. But I found that there was a need for customers and distributors to use this chatbot as well from questions asked online in our developer community. \cite{DEV} Accordingly, I changed the model of the chatbot from RAG to RAG-Fusion with the hypothesis that it would provide answers that not only held the same accuracy as the RAG chatbot, but were also more comprehensive in addressing various perspectives of users' questions. Thus, I tested the bot for viability in the following three areas: answering product-specific questions from engineers, answering sales teams' queries about sales strategies, and answering customer queries about products. Specifically, I focused on answering questions related to our silicon MEMs microphones  and metal-oxide-semiconductor field-effect transistors (MOSFETs). \cite{MEMS} \cite{MOSFETS}

\section*{RAG vs RAG-Fusion}
\label{sec:headings}

The traditional retrieval-augmented generation chatbot model for specific product information consists of the following steps:
\begin{itemize}
\item Gather product documents (e.g., datasheets, and product selection guides) in a database of documents for retrieval into text.
\item Create vector embeddings -- numerical representations of the text that the algorithm can understand -- and store them in a vector database.
\end{itemize}\vspace*{3pt}
Upon receiving a query from the user,
\begin{itemize}
\item Retrieve the $n$ most relevant documents based on vector distance to the original query via vector search.
\item Send the query together with the retrieved documents to a large language model to generate a response and output the response.
\end{itemize}\vspace*{3pt}

RAG-Fusion, on the other hand, has a few extra steps. \cite{RAGGIT} Once the original query is received, the model sends the original query to the large language model to generate a number of new search queries based on the original query. For example, if the user's original query is "Tell me about MEMs microphones," the generated queries may include
\begin{itemize}
\item \textit{What are MEMs microphones and how do they work?}
\item \textit{What are the advantages of using MEMs microphones?}
\item \textit{What are some recommended MEMs microphones?}
\end{itemize}

The algorithm then performs vector search to find a number of relevant documents like with RAG. But, instead of sending those documents with the queries to the large language model to generate the output, the model performs reciprocal rank fusion. Reciprocal rank fusion is an algorithm commonly used in search to assign scores to every document and rerank them according to the scores. The scores assigned to each document, or rrf scores, are \begin{equation} \text{rrfscore} = \frac{1}{\text{rank}+k} \end{equation} where rank is the current rank of the documents sorted by distance, and \textit{k} is a constant smoothing factor that determines the weight given to the existing ranks.
Upon each calculation of the score, the rrf score is accumulated with previous scores for the same document, and when all scores are accumulated, documents are fused together and reranked according to their scores. The model then sends the reranked results along with the generated queries and the original queries to the large language model to produce an output.

\begin{figure}[H]
  \centering
  \includegraphics[width=18.5pc]{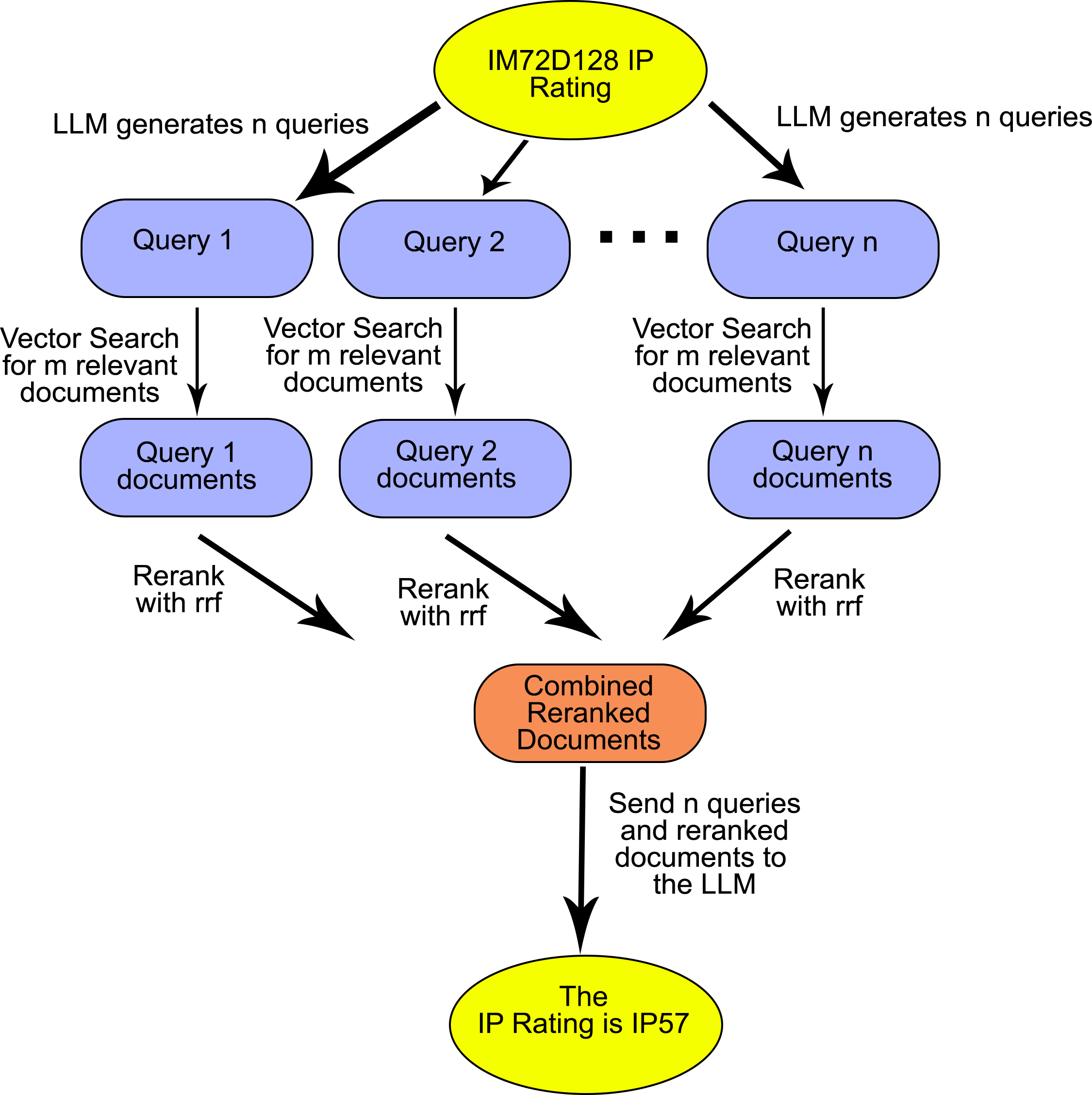}
  \caption{Diagram illustrating the high level process of RAG-Fusion starting with the original query ''IM72D128 IP Rating''}
  \label{fig:Diag1}
\end{figure}

\section*{Infineon Chatbot for Engineers}
There are three potential use cases of the Infineon RAG-Fusion chatbot: providing technical information to engineers, providing sales-oriented information to account managers, and providing customer-facing information to customers. 

To test the viability of the chatbot in providing information to engineers, I gathered a list of technical questions from Infineon's developer community forum regarding our products. \cite{DEV} I then optimized the questions to produce the best output with the RAG-Fusion. This included correcting spelling and grammar, converting statements into questions, and dividing multi-part questions into a single query. I then compared the accuracy of the RAG-Fusion chatbot in answering the technical questions against the correct forum solutions provided by Infineon experts. Take the following question for example.

\begin{quote}
\textbf{Query:} IP rating of mounted IM72D128 \newline
\textbf{Generated queries:} ['1. What is the IP rating of the mounted IM72D128?', '2. IP rating explained for mounted IM72D128.', '3. Waterproofing capabilities of the IM72D128 with its IP rating.', '4. How does the IP rating of the IM72D128 affect its durability when mounted?'] \newline
\textbf{Bot answer:} The mounted IM72D128 has an IP57 rating, indicating a high level of protection against both dust and water. The first digit, 5, signifies its resistance to solid objects like dust or dirt. The second digit, 7, represents its resistance to water. This rating means that the IM72D128 has a sealed design, preventing water or dust from entering between the membrane and backplate. This robust design ensures no mechanical blockages or electric leakage issues typically seen in MEMS microphones. As a result, the IM72D128 is highly durable and can withstand exposure to water and dust without significant damage. Its sealed MEMS design allows it to function properly even in challenging environmental conditions, providing long-term durability and reliable performance. \newline
\textbf{Forum solution: } Yes, you are right. IM72D128 has IP57 rating. The recommended Reflow soldering and board assembly of the microphone is given in section 7 (page: 14-15) of the datasheet. Please also refer to the application note: MEMS microphone mechanical \& acoustical implementation. \end{quote}

In this case, the chatbot provided a correct answer, that the IM72D128 microphone has an IP57 rating. Unlike the forum solution, however, the chatbot not only provided the rating, but also explained what the rating signifies. In addition, it provided information about the microphone's sealed design, ensuring no mechanical blockage or electric leakage issues, and a final statement summarizing its strengths. This extra information was derived from the generated queries and provided an answer that was not only correct, but also more comprehensive and had more context surrounding the question compared to the forum solution. The only part missing from the answer is that the microphone delivers high ingress IP57 protection at a microphone level because of Infineon's latest Sealed Dual Membrane MEMS technology. This missing part may be solved, however, by increasing the number of queries generated or documents retrieved.

In addition to technical product questions, many engineers also ask for help troubleshooting or product instructions. The bot, however, tends to answer these questions with very general instructions and by indicating the user should follow up with Infineon. The following query highlights this tendency.

\begin{quote} How to connect and communicate with the PDM interface of IM69D130 with stm32 \end{quote}
The bot lists six steps, five of which are mainly general guidance, with one step matching the expert solution:
\begin{quote} \textbf{Bot:} Configure the PDM interface settings on the STM32 microcontroller according to the requirements of the IM69D130, including clock frequency, data format, and other parameters. \newline
\textbf{Forum:} Set up the data format on the I2S interface to match the PDM data from the IM69D130. \end{quote}
However, the forum answer emphasizes that the IS2 interface of the microcontroller must be in slave receive mode, which is nowhere to be found in the bot's answer. This is most likely because the product database does not contain this kind of guidance. None of the documents mention slave receiver mode, which is a keyword in the response. Since it is unable to retrieve this information, the language model makes up for this by outputting its prior knowledge to supplement the missing information from the documents, which are general troubleshooting steps and not product-specific steps. This tendency can be solved by providing correct documentation in the database regarding troubleshooting and methods to use the product.

\section*{Infineon Chatbot for Account Managers}
Although its original purpose was to be used for engineers, the bot ended up performing the best in response to sales-oriented questions. These are questions that are asked by account managers or inside sales representatives that give insight into the benefits of Infineon products and sales strategies.

To devise sales strategies, the bot retrieves the documents related to the specific products, in the following case our 100V Linear Field-effect Transistor (FET). It then synthesizes the product knowledge with logic from the large language model to create a valuable strategy based on accurate product information. Generating multiple queries is key to devising a comprehensive and multiperspective sales strategy. For example, in response to the prompt \begin{quote} How do I sell a 100V OptiMOS Linear FET for power over ethernet to my customer? \end{quote} the bot generates unique prompts addressing market trends and customer preferences, best practices for selling electrical components, the value proposition of our Linear FET, and strategies for effectively pitching the benefits and features of our Linear FET. The answer generated by the chatbot addresses these queries by providing technical specifications such as the latest IEEE 802.3bt standard increasing the available power via Power over Ethernet (PoE). It continues to highlight Infineon's expertise in switched-mode power supply designs and our FETs' high efficiency, protection, best-in-class performance with ultra-low RDS(on) and low charge, and high reliability. It responds that the value proposition includes improving efficiency and power density, low conduction losses, and high switching performance leading to energy savings and better thermal management. For sales strategies, it suggests highlighting its value proposition, advantages in performance and efficiency, and the importance to understand customer-specific needs and requirements for their applications. It ends with suggesting the account manager provide technical support and assistance, which ties back into our Infineon chatbot use case for engineers.

This sales-oriented question answering for account managers reduces the need for an account manager to ask an engineer for product information as well as the engineer's need to sift through a 100-page datasheet. This saves times for everyone in the process and increases the account manager's ability to sell the product, resulting in more sales and higher revenue.

Inside sales representatives often need to quickly find product-specific information. This is similar to the use case for engineers, but instead of technical information like voltage and current density, it is information about applications and product validation.
\begin{quote}
\textbf{Query:} What is the application of IM73A135V01? \newline
\textbf{Bot response: } Based on the provided documents, the IM73A135V01 is a dust and water resistant analog XENSIVTM MEMS microphone with a best-in-class Signal to Noise Ratio (SNR) of 73dB(A) for high-quality audio pick-up. It also has a flat frequency response and low distortion. In addition, it is IP57 robust to dust and water. While the documents do not specifically mention application examples or industry applications for the IM73A135V01, it is a robust and high-performance MEMS microphone that can be used in various audio devices such as TWS earbuds and compact audio devices. Based on its features and benefits, it can be used in applications that require high-quality audio pick-up, low distortion, and dust/water resistance. Some potential applications could include consumer electronics, automotive, industrial, and smart devices. \newline
\end{quote}

\section*{Infineon Chatbot for Customers}
While the chatbot provides great support for sales teams, it is also able to provide comprehensive and helpful feedback for customers' product-related questions. These are not engineering questions, but rather questions from customers who are considering buying our products. For example, a customer could be looking to manufacture a camera and wants to know whether our microphones are a good fit for their camera. In fact, this was a real question asked on our developer community. \cite{DEV}
\begin{quote}
\textbf{Query:} Is the IM72D128 microphone good to use in an outdoor surveillance camera?
\newline
\textbf{Bot response:} The IM72D128V01 microphone is an ideal choice for outdoor surveillance cameras based on several documents. Document 1 states that it is IP57 dust and water resistant, making it able to withstand the conditions typically encountered in outdoor settings. Document 7 further describes it as an ultra-high-performance digital microphone with a high signal-to-noise ratio (SNR) and low distortions, which ensures clear and high-quality audio capture. This makes it well-suited for outdoor surveillance cameras that require excellent audio capture capabilities. Additionally, Document 8 mentions that the IM72D128V01 microphone is designed for both indoor and outdoor automotive applications, offers best-in-class audio performance, and has an extended temperature range. These features can be advantageous for outdoor surveillance cameras as well, emphasizing its suitability for capturing audio in outdoor environments.
\end{quote}
With this answer, the customer understands that our IM72D128 microphone does align with their project, resulting in a higher chance of a new sale.

\section*{Benefits of a Chatbot}
Customers and engineers tend to ask several questions in one query, and Infineon experts tend to respond to the main question and not the several smaller ones asked. With a chatbot, albeit in separate queries, users can ask all their questions and get answers.

In addition to product-specific knowledge, a chatbot can leverage pretrained knowledge from its large language model that an Infineon expert may not have. For example, one question asked by a customer was regarding whether MEMs microphones could be used in wooden instruments or professional recording studios. While our expert responded by pointing to the MEMs microphone product page and datasheets, the RAG-Fusion bot was able to provide a useful and comprehensive answer drawing on its prior trained knowledge and database of MEMs microphones.
\begin{quote}
MEMS (Micro-Electro-Mechanical Systems) microphones are small devices that convert sound waves into electrical signals. They utilize a diaphragm that vibrates in response to sound waves and an acoustic sensor that detects these vibrations. The sensor then translates the vibrations into electrical signals, which can be amplified and processed for various applications. \newline
The advantages of using MEMS microphones in stage or professional studio recordings are significant. Their small size allows for easy integration into equipment and placement in different areas. They offer high-performance audio capture with low self-noise, wide dynamic range, and high sensitivity. MEMS microphones also provide excellent audio quality, enabling accurate and precise sound reproduction. Additionally, their compact design enables the creation of microphone arrays for noise cancellation and beamforming, enhancing audio capture in complex acoustic environments. \newline
However, when using MEMS microphones in wooden instruments on stage or in professional studio recordings, there may be limitations or challenges. One potential limitation is their sensitivity to handling noise, as MEMS microphones can be more sensitive to vibrations and movements compared to traditional microphones. This requires careful handling and mounting to minimize unwanted noise. Additionally, the frequency response of MEMS microphones may not perfectly match the tonal characteristics of wooden instruments, necessitating careful selection and placement to achieve the desired sound. \newline
To capture the sound of wooden instruments effectively, some recommended MEMS microphones include models like the Infineon-IM73D122 or the Infineon-IM72D128. These microphones offer high sensitivity, low self-noise, and a wide dynamic range, making them suitable for capturing the nuances and details of wooden instruments. However, it is important to experiment and select the microphone that best suits the specific instrument and desired sound. \newline
In summary, MEMS microphones can be used in wooden instruments on stage or in professional studio recordings, offering advantages such as small size, high audio quality, and the ability to capture details and nuances. However, careful consideration should be given to microphone selection, placement, and handling to overcome any potential limitations or challenges.
\end{quote}

\section*{Challenges of RAG-Fusion}

The largest challenge of RAG-Fusion is the slow answer time from receiving the query to outputting the answer. I compared the runtime to our traditional RAG chatbot by performing ten back-to-back runs with the same query. I then subtracted the time when the query was received from the time the output was given to determine the time it took for that run. Back-to-back runs should control for APIs having different response times at different times of the day.

\begin{table}[H]
 \caption{Comparison of RAG-Fusion Time and RAG Time in Smart Speaker Runs}
  \centering
  \begin{tabular}{lll}
    \toprule
    Run     & RAG-Fusion Time (s) & RAG Time (s) \\
    \midrule
    1       & 42.72               & 30.48       \\
    2       & 32.05               & 32.93       \\
    3       & 12.85               & 25.94       \\
    4       & 42.78               & 16.70       \\
    5       & 36.58               & 11.89       \\
    6       & 45.99               & 10.62       \\
    7       & 34.92               & 17.58       \\
    8       & 35.56               & 14.42       \\
    9       & 37.55               & 28.21       \\
    10      & 25.19               & 6.44        \\
    \textbf{Average} & \textbf{34.62} & \textbf{19.52} \\
    \multicolumn{3}{l}{\textbf{Observation:} RAG-Fusion takes 1.77 times longer.} \\
    \bottomrule
  \end{tabular}
  \label{tab:rag_comparison}
\end{table}

Although it is hard to generalize exact runtime numbers, this average over ten back-to-back runs shows that RAG-Fusion bot is almost certainly slower than the RAG bot. In this case, RAG's average runtime from query to output was 19.52 seconds, with RAG-Fusion being nearly 1.77 times as slow with an average query to output time of 34.62 seconds. Moreover, this trend is evident in all individual runs but runs two and three.

After performing several runs with different queries and at different times, I determined that the slowness of RAG-Fusion compared to RAG can be mostly attributed to the second API call to the large language model. Even in long queries of 70 or more words, the call to generate multiple queries never took more than 5 seconds. The model then ran through document retrieval and reciprocal rank fusion almost instantly until it stopped for the second API call for several seconds. The second API call's complexity is amplified by the volume and diversity of the input in the form of multiple queries and a substantial number of documents in contrast to the much simpler call in RAG with one query and fewer documents. Two ways to overcome this slowness included hosting an LLM locally to reduce latency in the calls to the LLM and decreasing the number of queries generated by the first API call.

Another challenge of RAG-Fusion is the inability to empirically evaluate answers. While evaluation frameworks like ROUGE, BLEU, BLEURT, and METEOR are often applied to assess the accuracy of retrieval-augmented generation models, in the case of the Infineon chatbot, there is not necessarily a direct, expected answer. Thus, such evaluation frameworks are less effective for tasks such as the sales and customer-oriented answering where outputs can vary significantly in structure and content while still being correct. 

New evaluation toolkits such as RAGElo and Ragas have recently emerged. Such frameworks seek to provide automated evaluation for RAG-based models. RAGElo takes in a list of questions, documents, and answers. It then assigns a tournament-style Elo ranking of multiple RAG pipelines over different queries and prompts \cite{RAGELO}. Ragas provides empirical assessment scores in context precision, faithfulness, and answer relevancy \cite{RAGAS}. Despite promising results on evaluating RAG-based agents, further tweaking to these evaluation methods is needed to align with the unique goals of the Infineon RAG-Fusion chatbot.

The evaluation method that reflects RAG-Fusion’s goals are human evaluations based on accuracy, relevance, and comprehensiveness. Ratings from Infineon account managers and engineers provide the best insight into answer quality from an expert lens. For the purposes of this paper, however, I manually evaluated the responses based on the same scale. I found that RAG-Fusion excels in accuracy and comprehensiveness, but occasionally strays in relevance. This occurs when irrelevant queries are generated by the large language model. While the answers to the generated queries were accurate and comprehensive, they strayed from the intention of the original question if the generated query it answered differed too much from the intent of the original query.

To achieve answers that completely match the intention of the original query, users must sometimes engineer prompts to be more specific and comprehensible so that the large language model generates appropriate queries. An example question from a distributor was regarding whether our IM72D128 microphone is good for them to use in a security camera: "If it was  Okay, it is very good for out-door surveillance camera," with "If it was Okay" referring to whether water and dust come through the sound port. The first prompt I tried for this question was "Is IM72D128 very good for an out-door surveillance camera?" However, the language model did not recognize my intention of the phrase "good for" and thus generated prompts labeling IM72D128 as a surveillance camera itself. I was able to generate relevant queries with IM72D128 understood to be a microphone used in a surveillance camera with the prompt "Is the IM72D128 microphone good to use in an outdoor surveillance camera?" Situations like this can be remedied by providing a prompt creation guide to users, but in the long run I would like to investigate ways to integrate prompt engineering into the bot side rather than forcing the burden on users.

Another challenge of RAG-Fusion, which is also a challenge of RAG, is its inability to provide definitively negative answers. Below is an example of when the bot could not find documents relating to the user's query. 

\begin{quote}
\textbf{Query:} We need to Confirm whether the IFX waterproof MIC has a Sleeping mode and wake-up functions \newline
\textbf{Response:} Based on the provided documents, it is uncertain whether the IFX waterproof MIC has sleeping mode and wake-up functions, as none of the documents explicitly mention these features. To obtain accurate information, it is recommended to refer to the official documentation or reach out to the manufacturer directly for confirmation. \newline
\textbf{Forum solution:} Our microphone does not come with sleep mode and wake up functionality.
With respect power optimization, our microphone comes with low power mode.
\end{quote}

While the bot was correct that there is no mention of sleeping mode and wake-up functions in the database of product information, it did not provide a firmly negative answer. This is because the nature of the bot is centered around document retrieval and not providing clear negative answers.

\section*{Conclusion}
I found that the Infineon RAG-Fusion chatbot was able to provide more accurate and comprehensive answers than traditional RAG models. This includes answers for questions related to technical product information for engineers, sales strategies for account managers, and customer-oriented product explanations. This is due to the generation of multiple queries based on the original query, which allows the chatbot to answer questions from multiple angles. In addition, using RRF to rerank the documents retrieved results in the most relevant documents getting the highest priority in answer generation. 

While RAG-Fusion increases answer quality, it comes with challenges such as a longer runtime than other models due to a more complex call to the LLM with multiple queries and more documents, answers going off track from the original query due to irrelevant queries generated by the first LLM call, and the occasional need for appropriate prompt engineering to generate the desired outcome.

Many questions asked on the developer forum were from customers and distributors in non-English speaking countries. When customers translate their original questions to English, some context may be lost and thus, customers may not receive the answers they want from Infineon experts. This was clear with the previously discussed question regarding the usage of MEMs microphones in security cameras. In the future, I will expand the Infineon RAG-Fusion chatbot to answer in languages other than English, especially Japanese and Mandarin Chinese. 

I will also research ways to better represent multimodal pdf datasheets as text for retrieval-augmented generation. This will allow RAG-Fusion chatbots to retrieve answers from more granular text, resulting in a higher accuracy of specific product information and reduced hallucination frequency. I will evaluate and tweak systematic frameworks like RAGElo and Ragas according to RAG-Fusion’s accuracy and performance. Other future research includes optimizing methods to improve real-time performance optimization, automated quality assurance, and integration into internal and external web platforms.

\section*{Acknowledgements}
The author would like to thank Brooks Felton, Cynthia Meah, and Christopher Arnold.

\section*{About the Author}
Zackary Rackauckas is an AI and business development research intern at Infineon Technologies, San Jose, United States. His current interests include multilingual NLP, generative AI, and digitization and decarbonization methods. He is a professional member of IEEE Intelligent Systems. Zack received his BA in mathematics from Swarthmore College. Contact him at zackary.rackauckas@infineon.com.

\vspace*{-8pt}

\end{document}